\documentclass[manuscript,screen]{acmart}
\usepackage{tabu}
\usepackage{tabularx}
\usepackage{amsmath}
\usepackage{bbm}
\usepackage{graphicx}
\usepackage{caption}
\usepackage{subcaption}
\usepackage{hyperref}
\usepackage{url}
\usepackage{cases}
\usepackage[english]{babel}
\usepackage{blindtext}
\usepackage{bm}
\usepackage{multirow}
\usepackage{makecell}
\usepackage{balance}

\usepackage{algorithm}
\usepackage{algpseudocode}

\renewcommand\footnotetextcopyrightpermission[1]{} %

\newenvironment{packed_itemize}{
\begin{list}{\labelitemi}{\leftmargin=1.5em}
  \setlength{\itemsep}{1pt}
  \setlength{\parskip}{0pt}
  \setlength{\parsep}{0pt}
  \setlength{\headsep}{0pt}
  \setlength{\topskip}{0pt}
  \setlength{\topmargin}{0pt}
  \setlength{\topsep}{0pt}
  \setlength{\partopsep}{0pt}
}{\end{list}}

\begin{document}


\title{BehaveGPT: A Foundation Model for Large-scale User Behavior Modeling}

\author{Jiahui Gong}
\affiliation{%
  \institution{Department of Electronic Engineering, BNRist, Tsinghua University}
  \country{Beijing, China}}
\email{gjh22@mails.tsinghua.edu.cn}

\author{Jingtao~Ding}
\affiliation{%
  \institution{Department of Electronic Engineering, BNRist, Tsinghua University}
  \country{Beijing, China}}
\email{dingjt15@tsinghua.org.cn}

\author{Fanjin Meng}
\affiliation{%
  \institution{Department of Electronic Engineering, BNRist, Tsinghua University}
  \country{Beijing, China}}
\email{mengfj23@mails.tsinghua.edu.cn}

\author{Chen Yang}
\affiliation{%
  \institution{Honor Device Co., Ltd	}
  \country{Beijing, China}}
\email{yangchen6@honor.com}

\author{Hong Chen}
\affiliation{%
  \institution{Honor Device Co., Ltd	}
  \country{Beijing, China}}
\email{chenhong3@honor.com}

\author{Zuojian Wang}
\affiliation{%
  \institution{Honor Device Co., Ltd	}
  \country{Beijing, China}}
\email{wangzuojian@honor.com}

\author{Haisheng Lu}
\affiliation{%
  \institution{Honor Device Co., Ltd	}
  \country{Beijing, China}}
\email{luhaisheng@honor.com}

\author{Yong Li}
\affiliation{%
  \institution{Department of Electronic Engineering, BNRist, Tsinghua University}
  \country{Beijing, China}}
\email{liyong07@tsinghua.edu.cn}

\renewcommand{\shortauthors}{Jiahui Gong et al.}

\begin{abstract}
In recent years, foundational models have revolutionized the fields of language and vision, demonstrating remarkable abilities in understanding and generating complex data; however, similar advances in user behavior modeling have been limited, largely due to the complexity of behavioral data and the challenges involved in capturing intricate temporal and contextual relationships in user activities. To address this, we propose BehaveGPT, a foundational model designed specifically for large-scale user behavior prediction. Leveraging transformer-based architecture and a novel pretraining paradigm, BehaveGPT is trained on vast user behavior datasets, allowing it to learn complex behavior patterns and support a range of downstream tasks, including next behavior prediction, long-term generation, and cross-domain adaptation. Our approach introduces the DRO-based pretraining paradigm tailored for user behavior data, which improves model generalization and transferability by equitably modeling both head and tail behaviors. Extensive experiments on real-world datasets demonstrate that BehaveGPT outperforms state-of-the-art baselines, achieving more than than a 10\% improvement in macro and weighted recall, showcasing its ability to effectively capture and predict user behavior. Furthermore, we measure the scaling law in the user behavior domain for the first time on the Honor dataset, providing insights into how model performance scales with increased data and parameter sizes.

\end{abstract}


\maketitle
\begingroup\renewcommand\thefootnote{\textsection}

\section{Introduction}

\begin{figure}[tb]
    \centering
    \includegraphics[width = 0.95\linewidth]{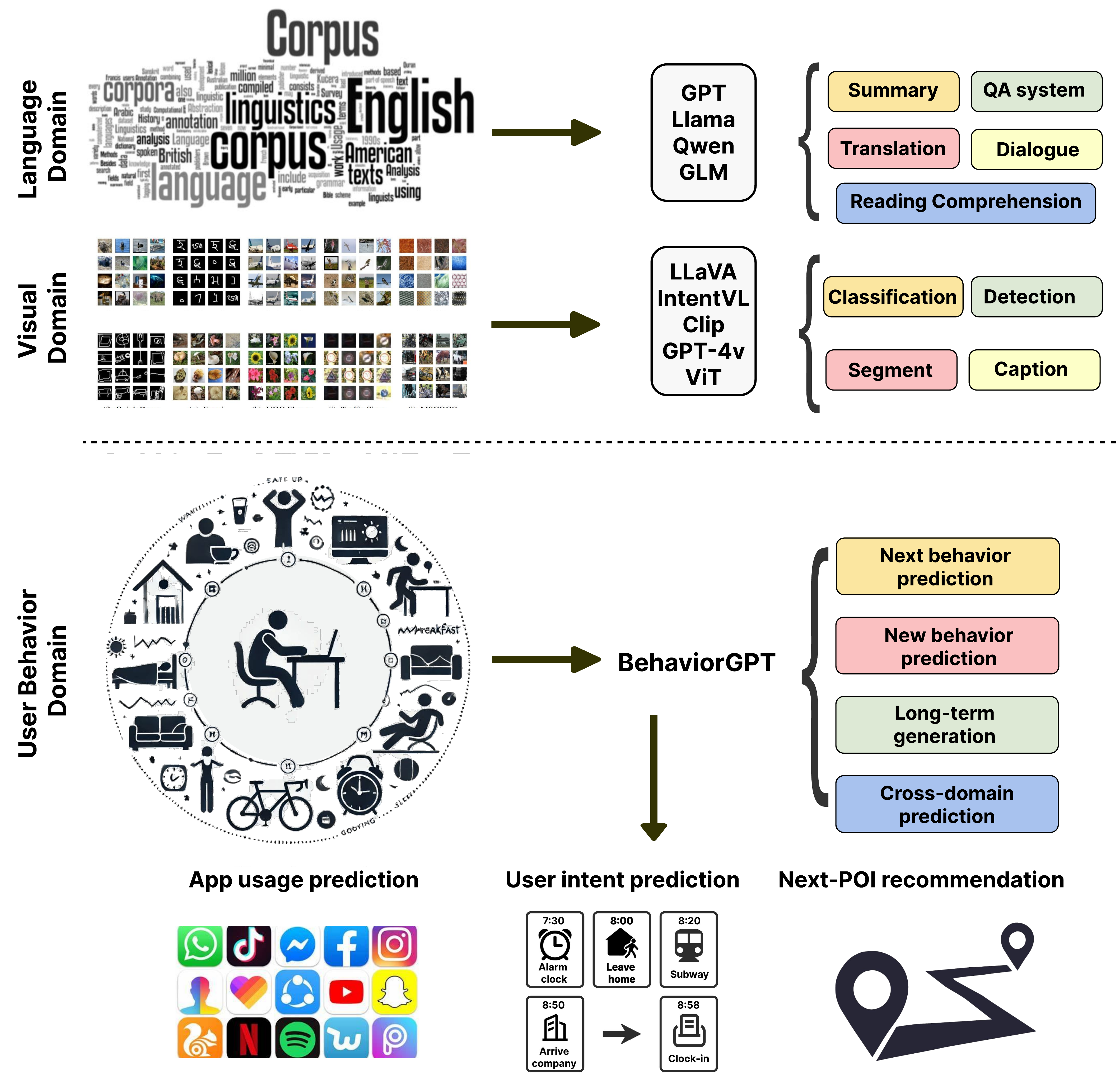}

    \caption{The paradigm of foundation model.}
    \label{fig:paradigm}
\end{figure} 

Recently, many foundational models have emerged in the language and vision domains \cite{gpt1,llama2,vit,mae}. These models are trained on massive datasets and exhibit strong generalization capabilities, allowing them to address various downstream tasks in their respective domains. However, these models primarily fall under the category of perceptual intelligence, which is considered an early stage of artificial intelligence \cite{morris2023levels, bandura1977social}. While perceptual intelligence enables models to perceive and process data, it has limitations when it comes to understanding, predicting user behaviors, and making decisions based on behavioral data, which are the key components of behavioral intelligence \cite{nadkarni2016superforecasting, hohwy2013predictive}. Behavioral intelligence is crucial for intelligent agents, such as robots \cite{tatarian2022does,yu2020rein}, automated systems \cite{steels2018building,jiang2022automatic}, and virtual characters \cite{ma2019exploring, pataranutaporn2021ai}, as it allows them to optimize their actions to achieve specific goals. These agents must be capable of perceiving their environment, learning from behaviors, and predicting future behaviors to adapt and make intelligent decisions.

User behavior prediction is a critical first step toward achieving behavioral intelligence, as it enables models to understand and predict future actions based on an individual’s historical behavior. This capability is essential for a wide range of applications, such as personalized recommendations \cite{li2023intent}, next point of interest (POI) predictions \cite{long2024diffusion}, and user intent predictions \ cite {pituning}. However, recent works \cite{llmesr,gao2023large} have primarily used large language models (LLMs) to model and predict user behavior. While LLMs excel at processing sequential data, they are inherently optimized for the language domain, where data distributions tend to be relatively balanced. In contrast, the behavior domain presents a starkly different challenge, characterized by highly imbalanced distributions: a small number of frequent behaviors dominate, while the majority consist of infrequent, long-tail behaviors. This imbalance complicates user behavior prediction, as LLMs often fail to capture the nuances, variations, and intricate dependencies inherent in diverse behavioral patterns. As a result, they struggle to model the full spectrum of user activities and provide equitable predictions for both common and rare behaviors.

In this paper, we propose BehaveGPT, a foundational model designed for user behavior modeling. Trained on vast amounts of user behavior data, more than 600 million behavior logs in total, BehaveGPT is capable of understanding and capturing diverse behavior patterns, leveraging its pretrained knowledge and flexible architecture to support a range of downstream tasks, such as next behavior prediction, new behavior prediction, long-term behavior generation, and cross-domain prediction. Specifically, we develop a transformer-based framework and introduce a new pretraining paradigm tailored to user behavior data, enabling more effective and fair modeling of user behaviors. This design enhances the model’s transferability and adaptation capabilities, making it more robust across different tasks and domains. Besides, we also explore the scaling phenomenon in the user behavior domain, offering insights into how model performance improves with increasing data and parameter sizes.
To summarize, our main contributions are as follows,

\begin{itemize}
    \item We are the first to train a foundational model specifically for the user behavior domain. After training on extensive user behavior data, we have developed a model that surpasses traditional LLMs in this area and supports multiple downstream tasks. Additionally, our initial exploration of the scaling phenomenon in the user behavior domain demonstrates that increasing both data size and model capacity significantly enhances performance, with models trained from scratch ultimately surpassing those initialized with pretrained LLM parameters as data scales.
    \item We propose a transformer-based framework and introduce a novel  DRO-based pretraining paradigm specifically tailored to user behavior data, allowing for more effective and equitable modeling of user behaviors to enhance the model’s transferability and generalization capabilities.
    \item Extensive experiments on three real-world datasets demonstrate the superiority of BehaveGPT over state-of-the-art baselines, achieving more than a 10\% improvement in macro and weighted recall and supporting various downstream tasks.
\end{itemize}

\section{Preliminary}
\subsection{Data Ananlysis}

We begin with a comprehensive data analysis, revealing a highly imbalanced distribution in user behavior data. A small number of common behaviors occur frequently, while the majority of behaviors are rarely observed. 

Figure \ref{fig:dis} illustrates the behavior distribution across two datasets, categorizing behaviors into four primary categories. The Honor dataset collects users’ mobile phone logs, while the Tencent dataset collects users’ trajectory data. As shown in the figure, the long-tail problem is apparent, and different behaviors dominate in each dataset. These variations highlight distinct user behavior patterns across different domains. 
Such an imbalance requires innovative solutions to ensure fair representation, making it essential to address both common and rare behaviors equitably to improve model robustness, generalization, and predictive accuracy across the entire spectrum of user activities.

\begin{figure}[t]
\centering
\subcaptionbox{Honor Dataset.\label{fig:dis_honor}}{
    \includegraphics[width=0.8\linewidth]{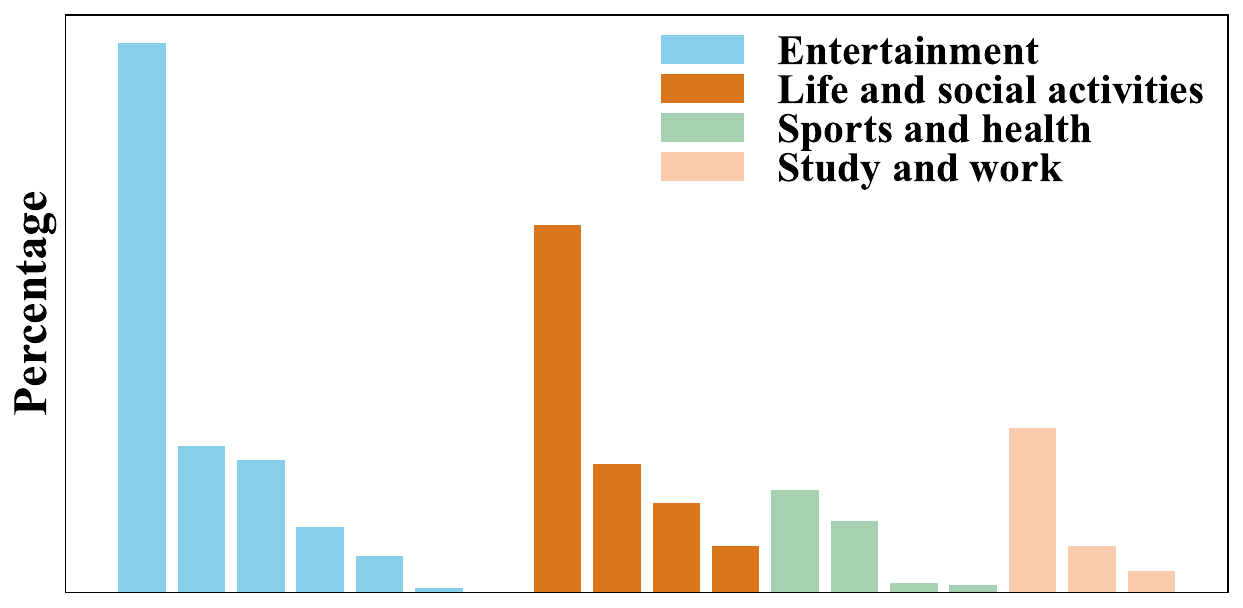}}
\subcaptionbox{Tencent Dataset.\label{fig:dis_mobile}}{
    \includegraphics[width=0.8\linewidth]{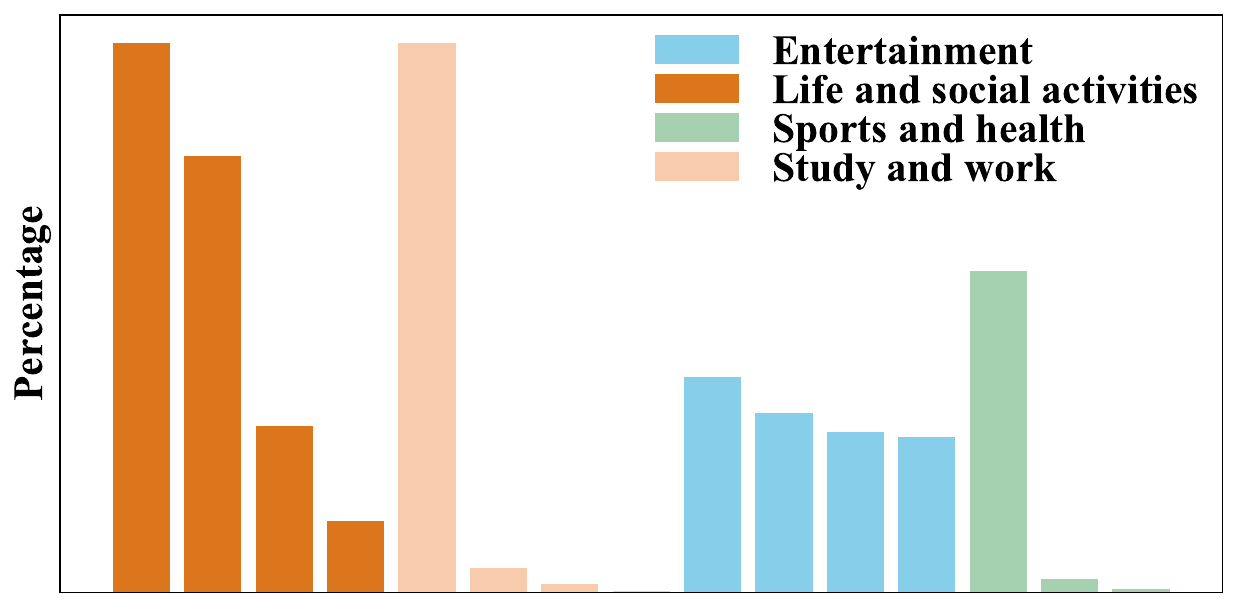}}
\caption{Distribution of behaviors in different datasets.}
\label{fig:dis}
\end{figure}

\subsection{Problem Definition}
Now we give a formal definition of our research problem:

\textsc{Definition} (Next behavior prediction). The user behavior can be represented as $x_i = \left(d_i, t_i, l_i, e_i \right)$, where $e_i$ denotes a specific event occurring at location $l_i$ during time slot $t_i$ on day $d_i$. Here, $d_i$, $t_i$, $l_i$, $e_i$, correspond to the weekday ID, time slot ID, location ID, and event ID, respectively. We use $\mathcal{D, T, L, E}$ to denote the sets of weekdays, time slots, locations, and events, with their respective sizes given by $N_D, N_T, N_L,$ and $N_E$. As outlined in the introduction, each user exhibits a particular behavior $b_i$ associated with an event-related behavior $x_i$. We define $\mathcal{B}$ as the set of possible behaviors, with its size represented by $N_B$.

The event encompasses specific instances involving users, such as the use of app services, spatial trajectory occurrences, and system-related events. While events represent all user-system interactions objectively, user behavior focuses on capturing the subjective actions, decisions, and responses that users exhibit during these events, reflecting individual tendencies or preferences. Therefore, the number of distinct user behaviors $N_B$ is typically less than the total number of events $N_E$.

While recommendation systems and next behavior prediction share the common goal of leveraging historical data to make predictions, they differ significantly in their objectives and scope. Recommendation systems aim to identify items most relevant to a user's preferences, typically using collaborative or content-based filtering techniques to optimize user engagement. In contrast, next behavior prediction seeks to forecast a user's specific future actions or behaviors based on sequential event data. This requires modeling complex temporal and contextual relationships within user activities, capturing not only preferences but also the intricate patterns and dependencies that drive behavior over time.
 The next behavior prediction task can be formed as,
\begin{equation}
\hat{b}_{t} = f(x_{t-I}, x_{t-I+1}, ..., x_{t-1})    
\end{equation}
 where $I$ denotes the length of the input size.

\section{Method}

\begin{figure*}[tb]
    \centering
    \subcaptionbox{The BehaveGPT architecture.\label{fig:dis_honor}}{
    \includegraphics[width=0.8\linewidth]{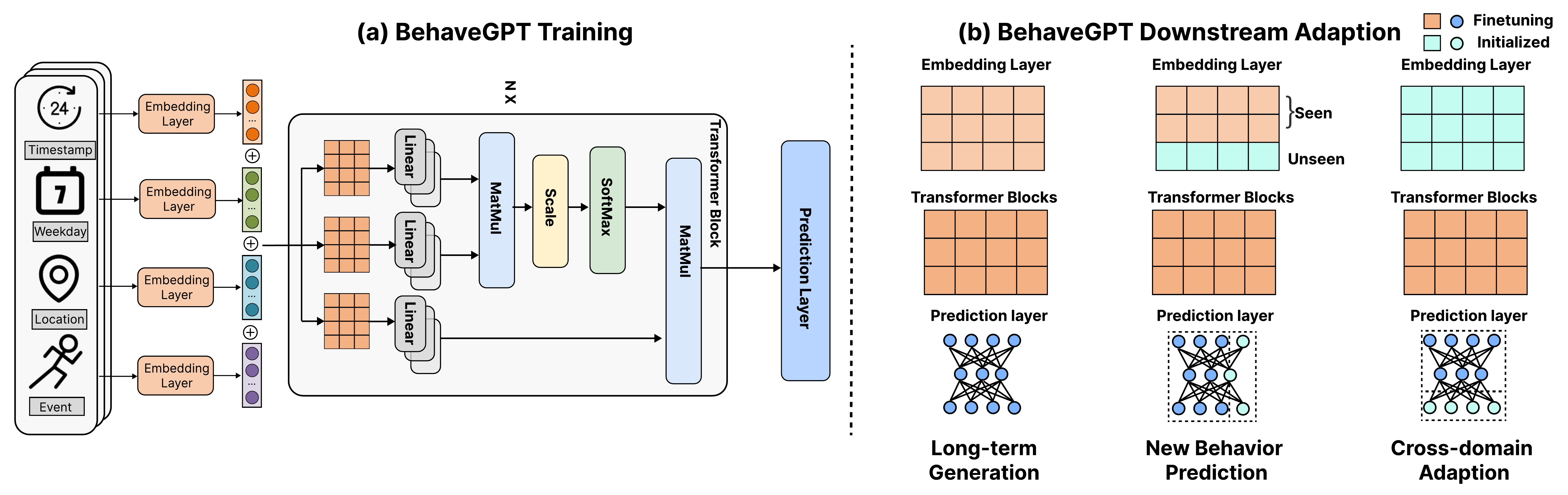}}
\subcaptionbox{The illustration of downstream task adaptation.\label{fig:dis_mobile}}{
    \includegraphics[width=0.8\linewidth]{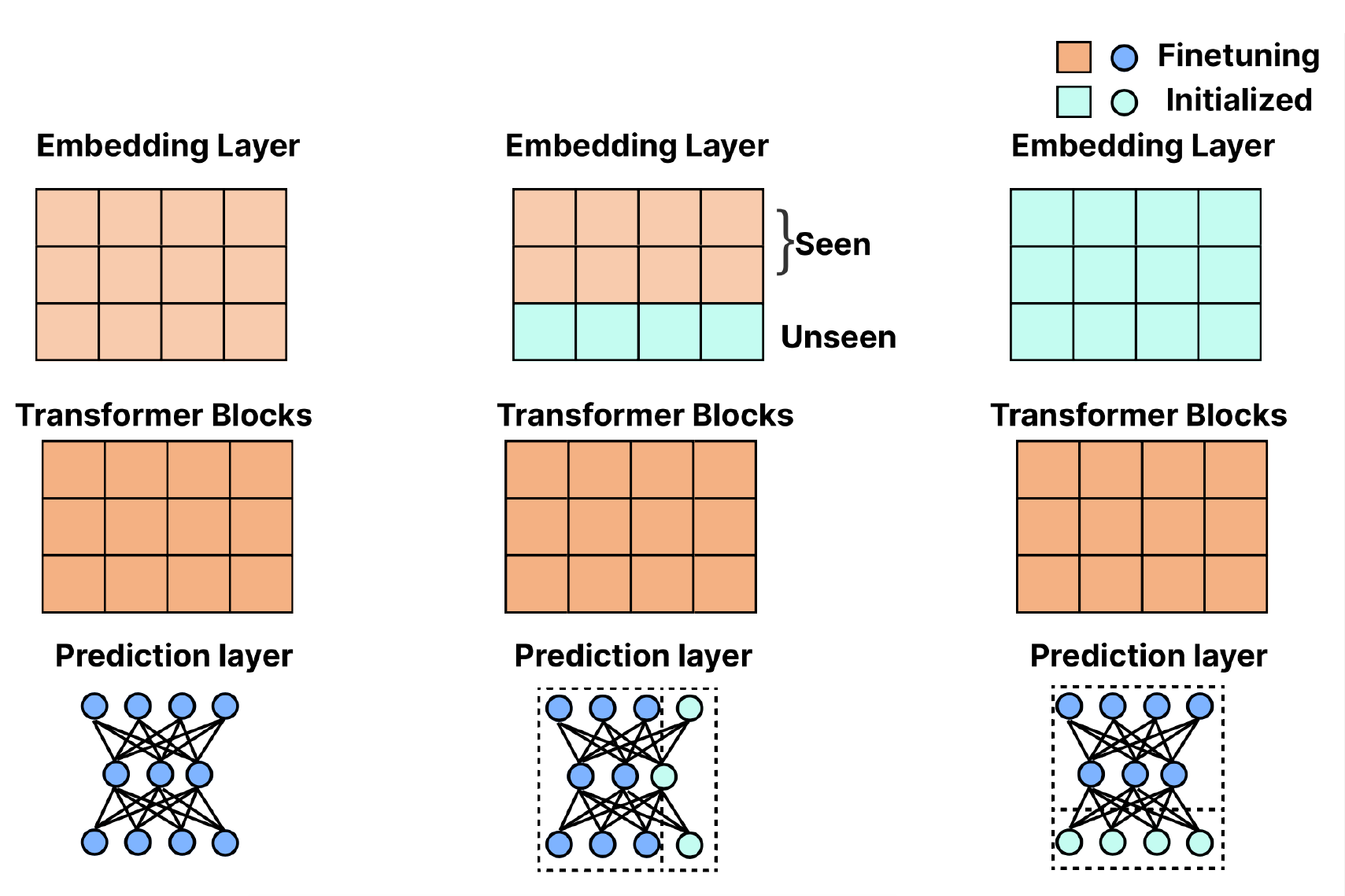}}
    \caption{The BehaveGPT framework.}
    \label{fig:framework}
\end{figure*} 

We introduce the BehaveGPT framework in Section \ref{sec:Foundation_Model_Architecture}, as shown in Figure \ref{fig:framework}. Input features, including weekday, timestamp, location, and historical events, are processed through embedding layers and transformer blocks to capture feature relationships. The prediction layer uses this information to forecast future user behavior. Section \ref{sec:Pretraining_Paradigm} details the DRO-based pretraining paradigm designed to minimize losses from uncertain tail-class distributions, ensuring fair modeling and improved generalization. Additionally, Section \ref{sec:Downstream_task_adaption} explains how the pretrained model adapts to various downstream tasks.

\subsection{Foundation Model Architecture}
\label{sec:Foundation_Model_Architecture}
BehaveGPT addresses user behavior complexities by capturing temporal, contextual, and spatial relationships. It encodes diverse features via embedding layers, learns dependencies with transformer blocks and self-attention, optimizes efficiency with Flash Attention, and predicts behaviors through an MLP layer.

$\bullet$ \textbf{Embedding Layer}
Since the input features cover the different aspect, we create four embedding layers to get the weekday embedding $\textbf{E}_w \in \mathbb{R} ^{I \times d}$, time slot embedding $\textbf{E}_t \in \mathbb{R} ^{I \times d}$, location embedding $\textbf{E}_l \in \mathbb{R} ^{I \times d}$ and historical event embedding $\textbf{E}_e \in \mathbb{R} ^{I \times d}$, respectively, where $d$ denotes the embedding size.

$\bullet$ \textbf{Transformer Blocks}
The transformer block consists of a stack of $N$ identical interaction layers, where each layer captures increasingly complex feature interactions by leveraging the transformer mechanism. This progressive stacking allows the model to learn higher-order dependencies between the input features to obtain an implicit representation of the historical event sequence $H_t \in \mathbb{R} ^{I \times 4d}$, which can be formed as,

\begin{equation}\label{equ:TransBlock}
    \textbf{H}_t = \text{Transformer}(\text{concat}(\textbf{E}_l, \textbf{E}_w, \textbf{E}_t, \textbf{E}_e)). 
\end{equation}

However, the vanilla transformer architecture has both computational and memory complexities of $O(I^2)$, where $I$ is the sequence length. As the sequence length increases, the complexity grows significantly, making it challenging for the transformer to handle long sequences, especially when multiple layers are stacked. To address this issue, We employ Flash Attention \cite{flashatten}, a novel attention mechanism that is significantly faster and more memory-efficient, enabling the model to process longer sequences with greater effectiveness. Flash Attention operates 2–4 times faster than standard attention mechanisms while reducing memory usage by 5–20 times, making it highly suitable for handling complex, large-scale behavioral data.

$\bullet$ \textbf{Prediction Layer}
We use a Multilayer Perception (MLP) to be the prediction layer, which can be formed as,
\begin{equation}\label{equ:MLP}
\textbf{m} =  \mathbf{W}_2(\sigma (\mathbf{W}_1\textbf{H}_t+\varepsilon _1))+\varepsilon _2,
\end{equation}
where $\mathbf{W}, {\varepsilon }$ are the trainable weight matrix and the bias matrix. The output of the MLP $\textbf{m}$ is the predicted user behavior distribution.

\subsection{DRO-based Pretraining Paradigm}
\label{sec:Pretraining_Paradigm}

\begin{figure}[tb]
    \centering
    \includegraphics[width = 0.95\linewidth]{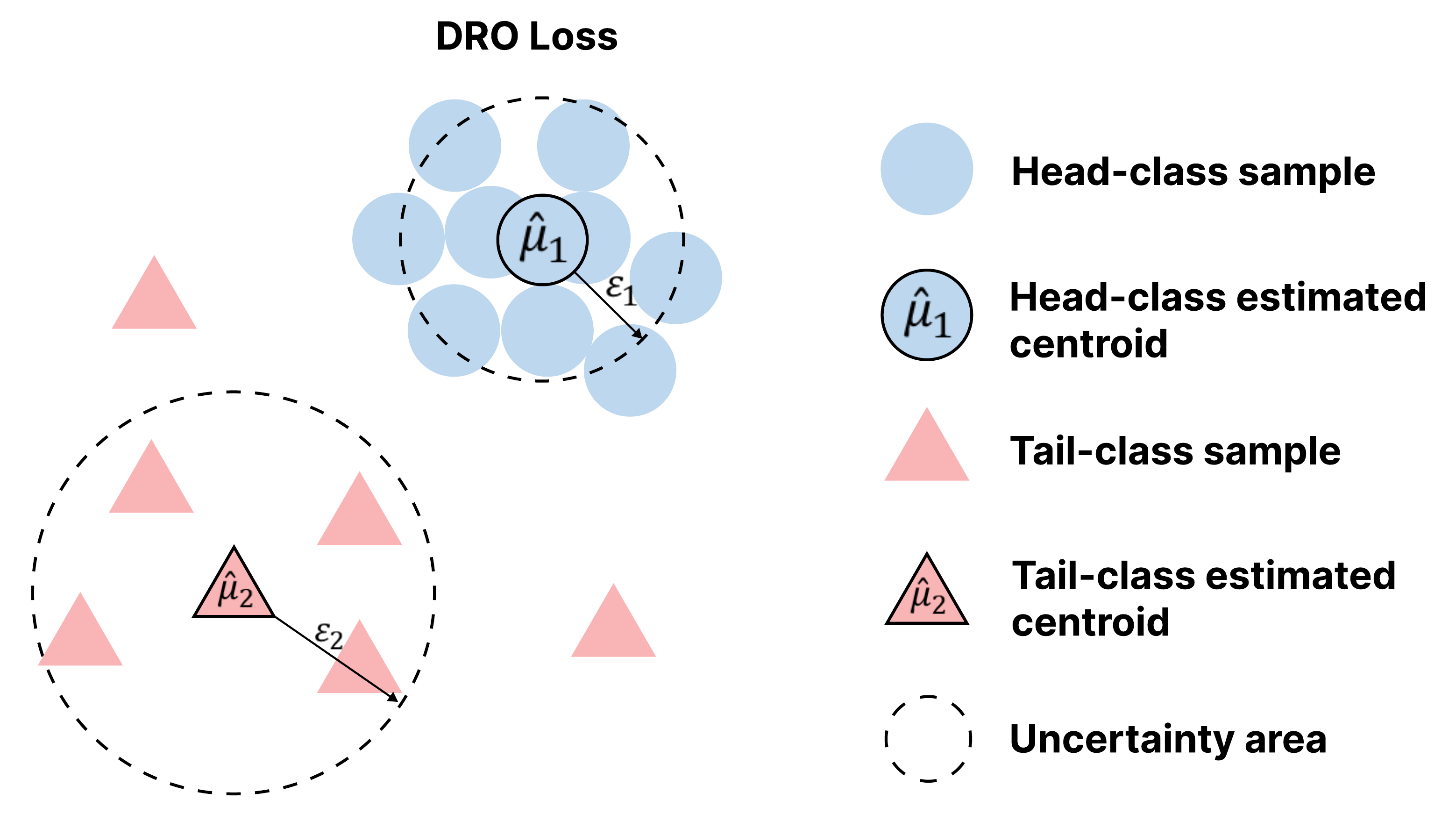}
    \caption{The illustration of DRO-based pretrain paradigm.}
    \label{fig:new_behavior}
\end{figure} 

The standard cross-entropy loss is widely used in classification tasks, but falls short when dealing with imbalanced data, such as user behavior datasets. It tends to prioritize head behaviors, which dominate the data distribution, while underperforming on long-tail behaviors with limited occurrences. This imbalance leads to biased predictions, where mainstream behaviors are overly emphasized, and rare behaviors are inadequately captured, resulting in suboptimal generalization across the behavior spectrum \cite{ding2020simplify}. To address this, we introduced the DRO-based pretraining paradigm specifically tailored to user behavior data to achieve fair modeling of user behaviors and improve the model's generalization ability.  Specifically, we incorporate distributional robust optimization (DRO)  \cite{dro}, an optimization method designed for tasks with inherent uncertainty, which allows for differences between the training and testing data distributions within a pre-defined uncertainty set, improving the model's resilience to distribution shifts.
As illustrated in Figure \ref{fig:new_behavior}, DRO defines a set of probability distributions $ p_b^\epsilon$ $, controlled by a scaling factor \epsilon$, which ensures that the model’s learned distribution does not deviate significantly from the training data distribution $p_b^{train}$. DRO seeks to minimize the worst-case expected loss by identifying the distribution within this set that maximizes the loss, thereby enhancing the model’s robustness to uncertainties. This formulation can be expressed as follows,
\begin{equation}
\label{equ:dro}
\begin{aligned}
    P_b^\epsilon :=  \{ p_b = & \epsilon p_b(b) \leq p_b^{train}(b) \ \forall b \} ,\\
    \sup_{p_b \in P_b^\epsilon} & \mathbb{E}_{b \sim p_b} \left[\ell(b; \theta)\right],
\end{aligned}
\end{equation}
where the $\ell$ denotes the cross-entropy loss. For the head-class behavior, they have abundant data, allowing the model to estimate their scaling factor with high confidence, resulting in a smaller uncertainty set. In contrast, tail-class behaviors have fewer data, making it difficult for the model to accurately estimate, leading to a larger uncertainty set.

This pretraining paradigm focuses on minimizing the potential loss from these uncertain tail-class distributions, ensuring the model performs well across both head and tail classes by accounting for the distribution of user behaviors, particularly improving its generalization on the user behavior domain.

\subsection{Downstream task adaption}
\label{sec:Downstream_task_adaption}
After training with a large amount of user behavior data, BehaveGPT offers a generalized representation and supports a range of downstream tasks, such as new behavior prediction, long-term generation, and cross-domain adaptation.

$\bullet$ \textbf{Next Behavior Prediction.}

Next behavior prediction is a core task in user behavior modeling and serves as the pretraining objective. By predicting a user's next behavior based on their historical behavior sequence, the model learns sequential dependencies and captures complex and diverse behavior patterns. This foundational task enhances the model’s performance across other downstream applications.

$\bullet$ \textbf{New Behavior Prediction.}

\textsc{Definition}: Behaviors that were not encountered during the pre-training phase can be predicted once they are introduced during the fine-tuning phase.

BehaveGPT excels in modeling previously unseen behaviors using only a small amount of new user behavior data. The model leverages its pretrained knowledge, advanced feature extraction capabilities, and deep contextual understanding to quickly adapt to new behaviors.

To adapt the model more quickly, we aim to transfer the parameters from the pretrained model to the greatest extent as shown in Figure \ref{fig:framework}. Specifically, for the embedding layer \( W_E \), previously encountered behavior embeddings are retained, while embeddings for new behaviors are initialized. Similarly, parameters in transformer blocks (\( W_Q, W_K, W_V \)) are fully transferred to ensure knowledge retention, facilitating rapid adaptation. In the prediction layer \( W_P \), weights for known behaviors are retained, and new weights are initialized for unseen behaviors, allowing efficient generalization with minimal fine-tuning. 

$\bullet$  \textbf{Long-term Generation.}

\textsc{Definition}: Input the user's historical behavior and predict the user's future $N$ behaviors.

BehaveGPT also supports long-term behavior generation by leveraging a user’s historical data. The model generates future behaviors in an autoregressive manner, iteratively predicting the next behavior and using it as input for subsequent steps. This enables the modeling of long-term behavioral trends and patterns, providing insights into user trajectories over extended periods.

$\bullet$  \textbf{Cross-domain Adaption.}

\textsc{Definition}: Transfer the knowledge learned from the pretrain dataset to the target dataset.

BehaveGPT demonstrates strong adaptability across domains after training on rich, diverse user behavior data. Unlike models like ChatGPT, whose embeddings are tailored for the language domain, behavior embeddings in BehaveGPT are more domain-specific and may not transfer seamlessly. To address this, we propose a cross-domain adaptation method that enhances the transferability of behavior embeddings. While embedding layers capture domain-specific representations, transformer blocks generalize effectively across domains by learning sequential dependencies and abstract patterns common to various behaviors. The MLP prediction layer also benefits from transfer learning, as intent prediction processes are often consistent across domains. However, the final projection layer is reinitialized to account for domain-specific differences in output mappings.

By reusing pretrained parameters from embedding layers and transformer blocks, the model requires fine-tuning of only a small subset of parameters to achieve excellent performance on new tasks or domains. This approach significantly reduces computational overhead while maintaining high adaptability and robustness, making BehaveGPT a practical and scalable solution for real-world behavior modeling challenges.

\section{Experiment}
\subsection{Experiment Settings}

\begin{table}[t]
\centering
\small
\caption{Statistics of the datasets used in our experiments. }
\scalebox{1.2}{
\begin{tabular}{ c| c c c} 
 \toprule
 \textbf{Datasets} & \textbf{Honor } & \textbf{Mobile } & \textbf{Tencent}  \\
 \midrule

 Type of Events & 114 & 2000 & 24435 \\
 Type of behaviors & 39 & 2000 & 14\\
 Users &81,102 & 1000 & 2102 \\
Duration & 3.1-5.31, 2024 & 4.20-26, 2016  &10.8-12.31, 2019\\
Number of logs & 205,605,167   & 4,171,950 & 463,437 \\

  \bottomrule
 
\end{tabular}}
\label{tab:datasets}
\end{table}

\subsubsection{Datasets.} 
We evaluate the performance of our model on three large-scale real-world behavior datasets.
\begin{packed_itemize}
\item \textbf{Honor Dataset.}  The Honor dataset, sampled from mobile phone usage logs, includes 81,102 users and over 2 million logs collected from March 1 to May 31, 2024. It comprises 114 event types categorized into 39 behaviors, such as news, education, work, and entertainment. The task involves predicting a user's next behavior based on their historical events.

\item \textbf{Mobile Dataset.} A publicly available check-in dataset with 4,171,950 app usage records from 1,000 users interacting with 2,000 apps between April 20 and April 26, 2016. Each record contains an anonymized app ID, location ID, and timestamp, with the goal of predicting the user's next app usage.  

\item \textbf{Tencent Dataset.}Sourced from social network mobility trajectories, this dataset includes 2,000 users and 463,437 records collected from October 8 to December 31, 2019. It features finely calibrated POIs annotated with specific intent types, aiming to predict users' next intents based on historical trajectories.  

\end{packed_itemize}
These three large-scale, fine-grained datasets comprehensively capture users' daily behavior activities, providing a rich foundation for modeling user behavior habits. By encompassing diverse dimensions such as app usage, mobility trajectories, and system events, these datasets enable a holistic understanding of user interactions and preferences.

\subsubsection{Baselines.}

We elaborately select the following ten representatives to be compared with our proposed algorithms, which cover large-scale recommendation methods (HSTU\cite{hstu}, Wukong\cite{wukong}), long-tail methods (LLM-ESR\cite{llmesr}, TailNet\cite{tailnet}, CDN\cite{cdn}, LIFT\cite{lift},  PITuning\cite{pituning}),  and sequential recommenders (SASRec\cite{sasrec}, PAT\cite{pat}, CASTR\cite{castr}). We provide the details of baselines in Appendix \ref{sec:baseline}.

\subsubsection{Metrics.}

To evaluate the model’s performance, we use widely adopted metrics \cite{pituning}. For prediction tasks, we measure the weighted precision ($Prec_w$), weighted recall ($Rec_w$), macro precision ($Prec_m$), macro recall ($Rec_m$), hit rate (HR) and NDCG(N). Weighted metrics and NDCG assess classification accuracy and ranking quality, while macro metrics evaluate average prediction accuracy across behaviors, highlighting performance for each behavior. For generation tasks, we use the K-S test (KS), Wasserstein distance (WD), Jensen-Shannon Divergence (JSD), BLEU, and Distinct-2, capturing distributional consistency, diversity, and alignment with target behaviors. Detailed calculations are provided in Appendix \ref{sec:metrics}.

\begin{table*}[t]
\centering
\caption{Overall Performance of Next Behavior Prediction. }

\scalebox{0.9}{
\begin{tabular}{c| c c c c c c | c c |c c c } 
 \toprule

& \multicolumn{6}{c}{\textbf{Honor Dataset}} & \multicolumn{2}{|c}{\textbf{Mobile Dataset}}& \multicolumn{3}{|c}{\textbf{Tencent Dataset}} \\ \midrule
     \textbf{Model} &$Prec_m$ &$Rec_m$ &$Prec_w$ &$Rec_w$&N@3&N@5&HR@3 &N@3 &$Rec_m$ &$Rec_w$&N@3\\ \midrule
    
    Wukong & 0.3101 & 0.2872 & 0.4809 & 0.4099 & 0.4961 & 0.6032 & 0.3596 & 0.2705 & 0.3201 & 0.5321 & 0.6385 \\ 
    HSTU & 0.1269 & 0.3396 & 0.4064 & 0.2486 & 0.3657 & 0.4126 & 0.2024 & 0.1455 & 0.1887 & 0.3655 & 0.5287  \\ 
    \midrule
    LLM-ESR & 0.3079 & 0.4688 & 0.4987 & 0.4143 & 0.5149 & 0.5983 & 0.3296 & 0.2312 & 0.3220 & 0.5621 & 0.6502 \\ 
    Tailnet & 0.3016 & 0.4657 & 0.5137 & 0.3609 & 0.5028 & 0.5503 & 0.2884 & 0.2153 & 0.3117 & 0.5378 & 0.6391\\ 
    CDN & 0.1644 & 0.3246 & 0.4854 & 0.3563 & 0.455 & 0.4912 & 0.2470 & 0.1931 & 0.1965 & 0.3828 & 0.5301  \\ 
    LIFT & \underline{0.3117} & \underline{0.4734} & 0.4963 & 0.3986 & 0.5352 & 0.6048 & 0.3388 & 0.2450 & \underline{0.3252} & 0.5658 & 0.6511  \\ 
    PITuning & 0.2883 & 0.4428 & 0.4989 & 0.4173 & 0.5206 & 0.5860 & 0.2700 & 0.2029 & 0.2878 & 0.5284 & 0.6330 \\ 
    \midrule
    PAT & 0.2243 & 0.3601 & \underline{0.5144} & \underline{0.4207} & \underline{0.5383} & \underline{0.6104} & \underline{0.3695} & \underline{0.297} & 0.3043 & \underline{0.5921} & \underline{0.6848} \\ 
    CASTR & 0.2219 & 0.3976 & 0.5036 & 0.3548 & 0.5001 & 0.5734 & 0.3078 & 0.2166 & 0.2368 & 0.3986 & 0.584 \\ 
    SASRec & 0.2144 & 0.2899 & 0.4587 & 0.3943 & 0.5128 & 0.5668 & 0.1772 & 0.1308 & 0.2090 & 0.3847 & 0.5370 \\ 
    \midrule
    BehaveGPT &\textbf{0.3729} & \textbf{0.5188} & \textbf{0.5709} & \textbf{0.4700} & \textbf{0.6140} & \textbf{0.6714} & \textbf{0.4140} & \textbf{0.3306} & \textbf{0.3507} & \textbf{0.6279} & \textbf{0.7431} \\ 
    Improv. & 19.63\% & 9.59\% & 10.98\% & 11.72\% & 14.06\% & 9.99\% & 12.04\% & 11.31\% & 7.84\% & 6.05\% & 8.51\% \\
  \bottomrule
\end{tabular}}
\label{tab:next_behavior_prediction}
\end{table*}

\subsubsection{Implementation Details.}
Our model employs the Adam optimizer with a learning rate of 0.0001. And we set the input length as 90. In the transformer block, the number of layers and dimension size are set as 12 and 1024, respectively. In the honor dataset, 70 days are allocated for training, 15 days for validation, and 15 days for testing. In the mobile dataset and trajectory dataset, 60\% of the data is used for training, with 10\% for validation and 30\% for testing.

\subsection{Next Behavior Prediction}

In the table \ref{tab:next_behavior_prediction}, we display the next behavior prediction result of BehaveGPT, with large-scale recommendations (HSTU, Wukong), long-tail methods (LLM-ESR, Tailnet, CDN, LIFT, PITuning) and sequential recommendation method (PAT, CASTR, SASRec) in three datasets. From the result, we have the following findings:

\begin{packed_itemize}
    \item \textbf{Our framework steadily achieves the best performance.} 
    Our model gets superior results on both datasets and performs better than other compared algorithms. For example, the macro metrics improvement of our model is around 7.84\% to 19.63\% compared with the second-best performance model (PAT). The $NDCG$ improvement of our model is about 8.51\% to 14.06\%.
    \item \textbf{Our model has the smallest difference between weighted metrics and macro metrics.} Our model achieves superior results across both macro and weighted metrics. While long-tail methods perform well on macro metrics, demonstrating their ability to improve predictions for long-tail behaviors, they do so at the cost of head behavior performance, leading to lower weighted metrics. On the other hand, sequential recommendation methods prioritize head behaviors, yielding higher performance on weighted metrics, but they tend to overlook long-tail behaviors, resulting in poorer macro metrics. 
\end{packed_itemize}

\subsection{New Behavior Prediction}

To evaluate the flexibility of our model, we conducted experiments by introducing a previously unseen user behavior and applying few-shot learning. we only used two weeks of data from 1,000 people as the finetuning dataset, which accounts for about 0.25\% of the pretraining dataset. and selected three behaviors with very different proportions and added them to the Honor dataset respectively, namely finance (5.22\%, high proportion), memo (1.04\%, medium proportion), and health query (0.04\%, low proportion). We then compared our model's performance against a meta-learning method, Mecos \cite{new_behavior_meta}, and an LLM-based finetuning method Recformer \cite{recformer}, where we use Llama3-8B as the backbone.

\begin{table}[t]
\centering
\caption{New Behavior Prediction Results on the Honor Dataset. }

\scalebox{1.2}{
\begin{tabular}{c|c c|c c|c c}
    \toprule
        model & Rec\_m & Finance & Rec\_m & Memo & Rec\_m & Health \\ \midrule
        Mecos & 0.3434 & 0.3673 & 0.3364 & 0.3109 & 0.3207 & 0.2935 \\ 
        Recformer & \underline{0.3849} & \underline{0.4000} & \underline{0.3858} & \underline{0.3388} & \underline{0.3969} & \underline{0.3523} \\ \midrule
        Our & \textbf{0.5076} & \textbf{0.642} & \textbf{0.491} & \textbf{0.6241} & \textbf{0.4847} & \textbf{0.6179} \\ 
        Improv. &31.88\% & 60.50\% & 27.27\% & 84.21\% & 22.12\% & 75.39\% \\ 
        \bottomrule
    \end{tabular}}
\label{tab:new_behavior_prediction}
\end{table}

\begin{table}[t]
\centering
\caption{Long-term generation result. }

\scalebox{1.2}{
\begin{tabular}{c|c c c c c}
    \toprule
        Model & KS & WD & JSD & BLEU & Distinct-n \\ \midrule
        SeqGAN &  \underline{0.1433}  &  \underline{1.0161}  &  \underline{0.1595}  &  \underline{0.2127}  &  \underline{0.3215}  \\ 
        DiffuSeq & 0.2547  & 2.0036  & 0.2282  & 0.1659  & 0.2943 \\ 
        UPC & 0.3562  & 1.9055  & 0.3187  & 0.1460  & 0.2710  \\ \midrule
        Our & \textbf{0.1286 } & \textbf{0.9597 } & \textbf{0.1327 } & \textbf{0.3228 } & \textbf{0.5326} \\ \bottomrule
    \end{tabular}}
\label{tab:generation}
\end{table}

\begin{table*}[t]
\centering
\caption{Cross-domain prediction result. }
\scalebox{0.95}{
\begin{tabular}{c | c c c c |c c c c c c } 
 \toprule

& \multicolumn{4}{|c}{\textbf{Mobile Dataset}}& \multicolumn{6}{|c}{\textbf{Tencent Dataset}} \\ \midrule
     \textbf{Model}&HR@3 &N@3&HR@5 &N@5 &$Prec_m$ &$Rec_m$ &$Prec_w$ &$Rec_w$&N@3&N@5\\ \midrule
        wukong & 0.3337 & 0.4141 & 0.262 & 0.3398 & 0.3836 & 0.3192 & 0.4557 & 0.5126 & 0.6153 & 0.6774 \\ 
        CASTR & 0.2530 & 0.3306 & 0.1945 & 0.3093 & 0.2988 & 0.2270 & 0.3704 & 0.3925 & 0.5708 & 0.6693 \\ 
        LIFT(GPT2-s) & \underline{0.3456} & \underline{0.4275} & \underline{0.2718} & \underline{0.3402} & \underline{0.4166} &\underline{ 0.3589} & \underline{0.5059} & \underline{0.5897} & \underline{0.6700} & \underline{0.7318} \\ \midrule
        Our & \textbf{0.4682} & \textbf{0.5623} & \textbf{0.3896} & \textbf{0.4304} & \textbf{0.5257} & \textbf{0.4696} & \textbf{0.6301} & \textbf{0.6941} & \textbf{0.7810} & \textbf{0.8090} \\
        Improv. &35.47\% & 31.53\% & 54.72\% & 26.51\% & 26.18\% & 30.84\% & 24.55\% & 17.69\% & 16.56\% & 10.54\%\\

  \bottomrule
\end{tabular}}
\label{tab:cross_domain_prediction}
\end{table*}

The new behavior prediction result, shown in Table \ref{tab:new_behavior_prediction}, reveals that our model has better flexibility in new behavior prediction, achieving 20\% improvement across all three new behaviors. While LLM-based fine-tuning methods demonstrate certain advantages in few-shot learning, they struggle to effectively model the complex patterns of user behavior and capture the intricate relationships between behaviors. This limitation hampers their overall performance in user behavior tasks. BehaveGPT, trained on large-scale behavior data, benefits from rich pretrained knowledge, strong feature extraction capabilities, and a deep understanding of context and semantics. This allows it to generalize effectively to new, unseen behaviors with minimal additional training.

\subsection{Long-term Generation}

To evaluate the model's generative capabilities, we conducted experiments by providing it with a historical behavior sequence and assessing the coherence and relevance of the generated sequences. For each sequence, the generation length was set to 50. We compared our method against SeqGAN \cite{yu2017seqgan}, DiffuSeq \cite{gong2022diffuseq}, and UPC \cite{liu2022upc}.

According to the results presented in Table \ref{tab:generation}, our model demonstrates exceptional performance by achieving the highest diversity and the closest alignment with real-world user behavior distributions. These results highlight the model's ability to generate user behavior sequences that not only exhibit a rich variety of patterns but also closely mirror the underlying structure and tendencies of the original data. The enhanced diversity reflects the model's capacity to capture a broader range of behavioral nuances, while the strong distributional alignment ensures that the generated sequences are both realistic and representative of real-world behaviors.

\subsection{Cross-domain prediction}
\subsubsection{Comparing different foundation models.}

To evaluate the cross-domain adaptability of our model after training on a large-scale user behavior dataset, we conducted experiments in which the model, initially trained on the Honor dataset, was adapted to the Mobile and Tencent datasets. We then compared its performance against several benchmarks, including the scaling recommendation method Wukong \cite{wukong}, cross-domain baselines CASTR \cite{castr}, and LLM-based fine-tuning methods LIFT, where we set the GPT2-small model as the backbone \cite{lift}.

The cross-domain results, shown in Table \ref{tab:cross_domain_prediction}, reveal that our model has better cross-domain ability, achieving more than 10\% improvement across all metrics. From the result, we have the following findings:
\begin{packed_itemize}
\item \textbf{The Honor datasets are more suitable for generalization to other behavior datasets.} The Honor dataset provides a comprehensive representation of diverse user behaivor patterns, while the traditional language models focusing primarily on the sequential structure of natural language, and recommendation systems aiming to identify items most relevant to a user's preferences, fail to capture the complex spatiotemporal relationships and behavioral diversity present in user behavior sequences, limiting their generalization capabilities. 
\item \textbf{Our model architecture is more flexible for cross-domain adaption.} 
BehaveGPT's model architecture and the DRO-based pretraining paradigm are specifically designed to capture hidden behavior patterns through its ability to model large-scale user behavior data. This flexibility allows the model to generalize effectively to various behavior domains by learning transferable representations that are robust across datasets.
\end{packed_itemize}
\begin{figure}[t]
\centering
\subcaptionbox{Mobile dataset.\label{fig:cross_method_mobile}}{
    \includegraphics[width=0.7\linewidth]{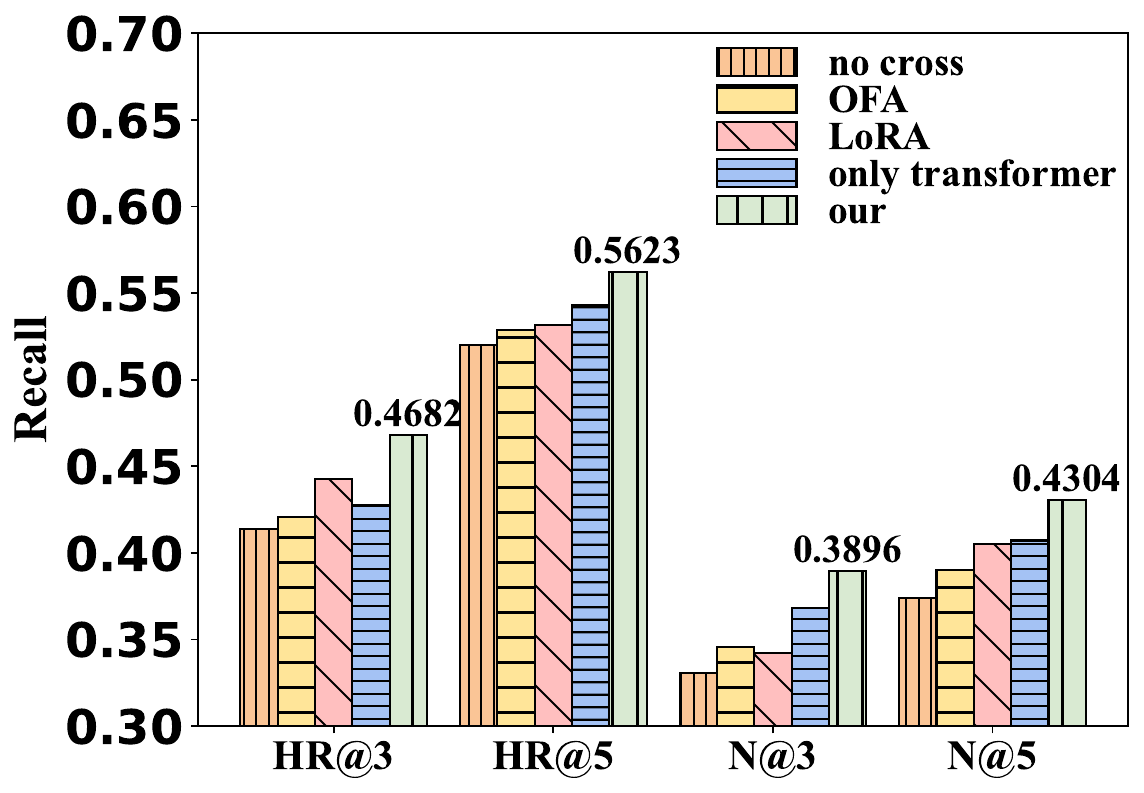}}
\subcaptionbox{Tencent dataset.\label{fig:cross_method_tencent}}{
    \includegraphics[width=0.7\linewidth]{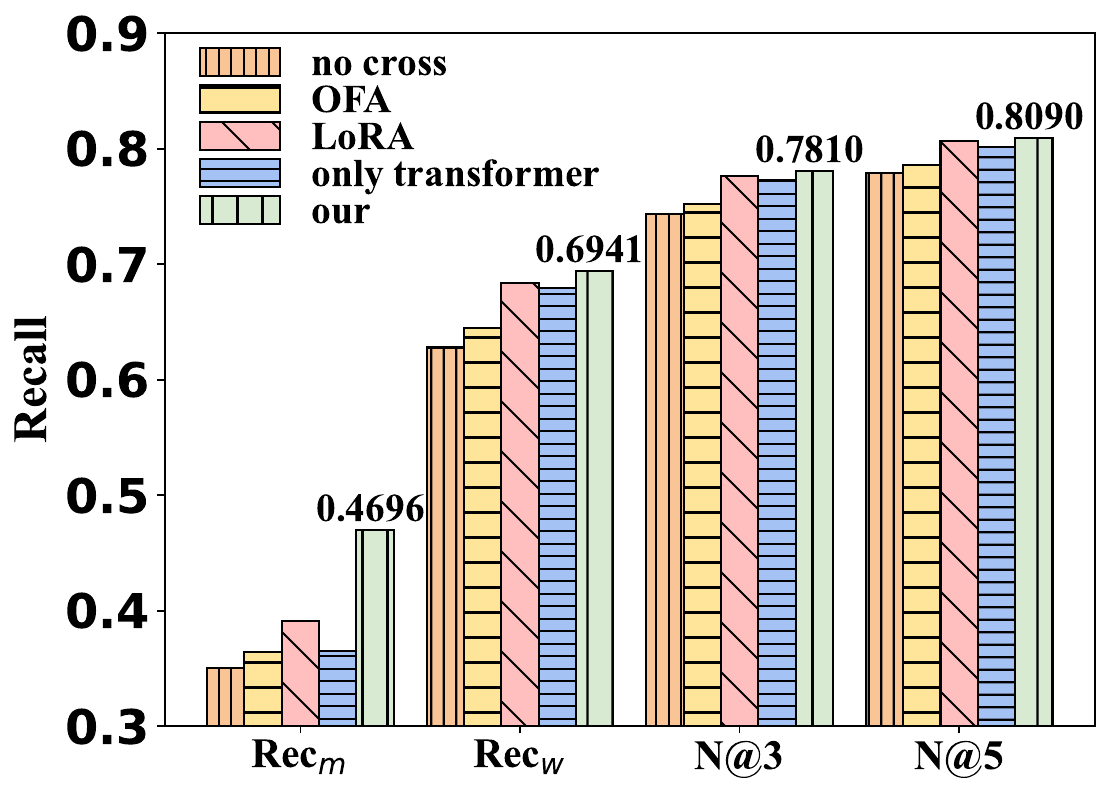}}
\caption{Comparison of cross-domain method.}
\label{fig:cross_method}
\end{figure}

\subsubsection{Comparing different adaptation techniques.}

Additionally, we evaluated the effectiveness of various adaptation techniques by comparing our approach with parameter-efficient methods, including LoRA \cite{lora}, OFA \cite{ofa}, and a method that transfers only the transformer blocks.

Figure \ref{fig:cross_method} compares various adaptation methods, showing that simultaneously transferring the parameters of both the transformer block and the prediction layer significantly improves the model's cross-domain performance. This comprehensive parameter transfer approach enhances the framework's adaptability and robustness across different domains.
Moreover, our framework is highly compatible with a wide range of parameter adaptation techniques, consistently delivering strong performance in diverse environments. By supporting various adaptation strategies, the model offers flexibility to meet specific requirements, whether related to computational constraints, accuracy goals, or the unique characteristics of the target domain. This versatility allows users to customize the adaptation process, ensuring optimal outcomes tailored to their particular application scenarios.

\begin{figure}[t]
\centering
\subcaptionbox{Long-tail behavior results in the Honor dataset.\label{fig:dro_honor}}{
    \includegraphics[width=0.95\linewidth]{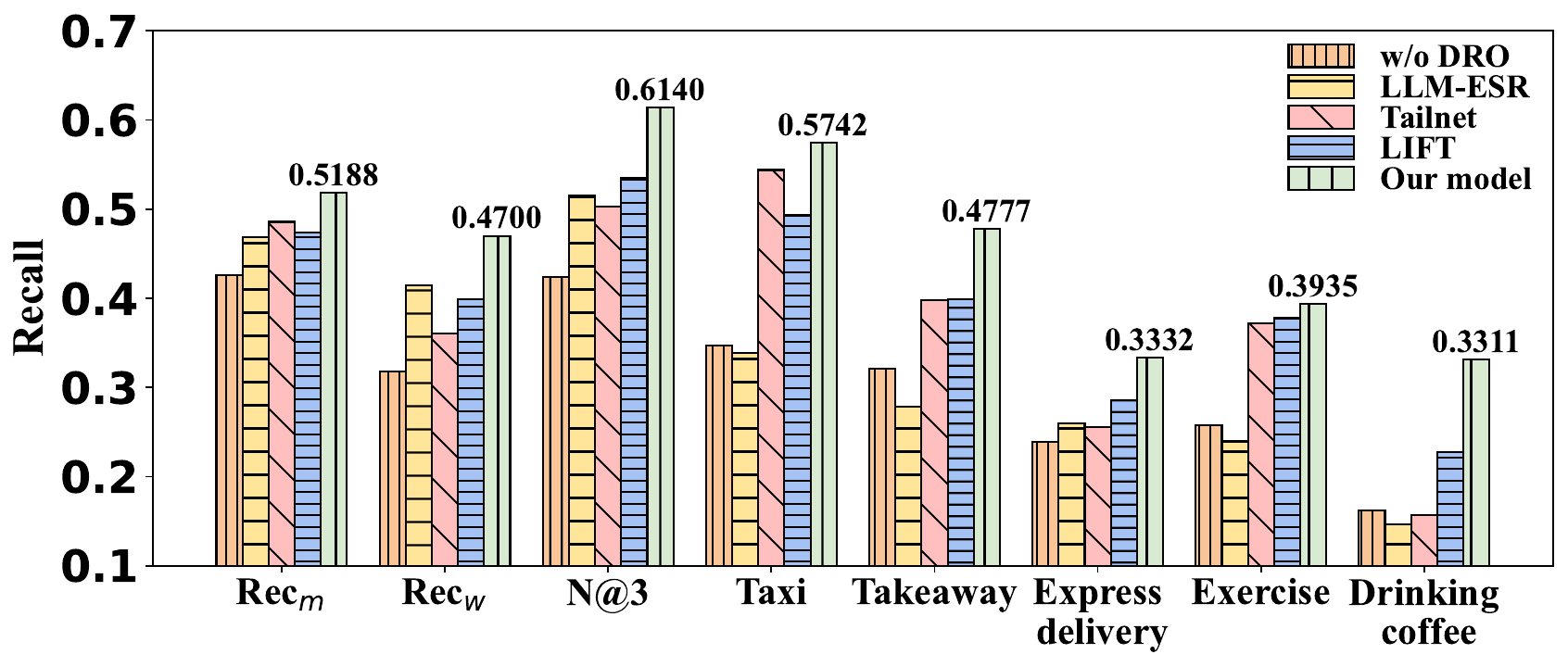}}
\subcaptionbox{Effectiveness of pertraining paradigm in cross-domain adaptation.\label{fig:dro_cross}}{
    \includegraphics[width=0.47\linewidth]{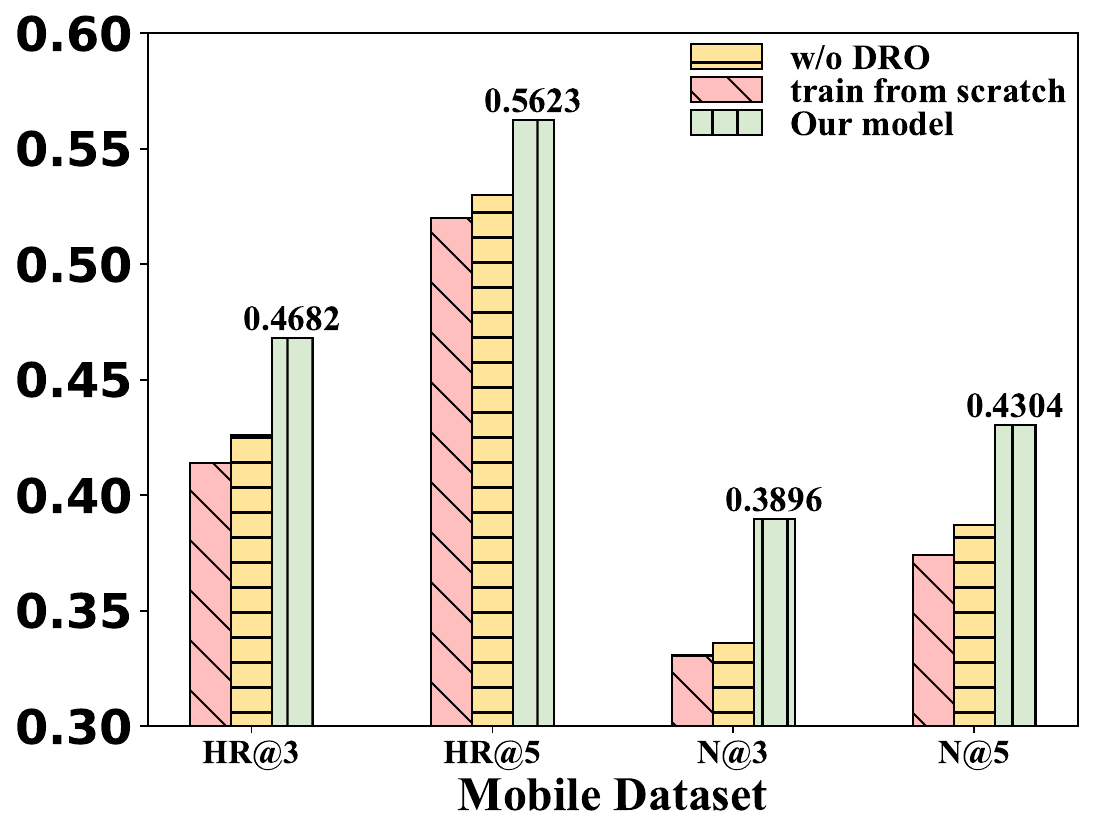}
     \includegraphics[width=0.47\linewidth]{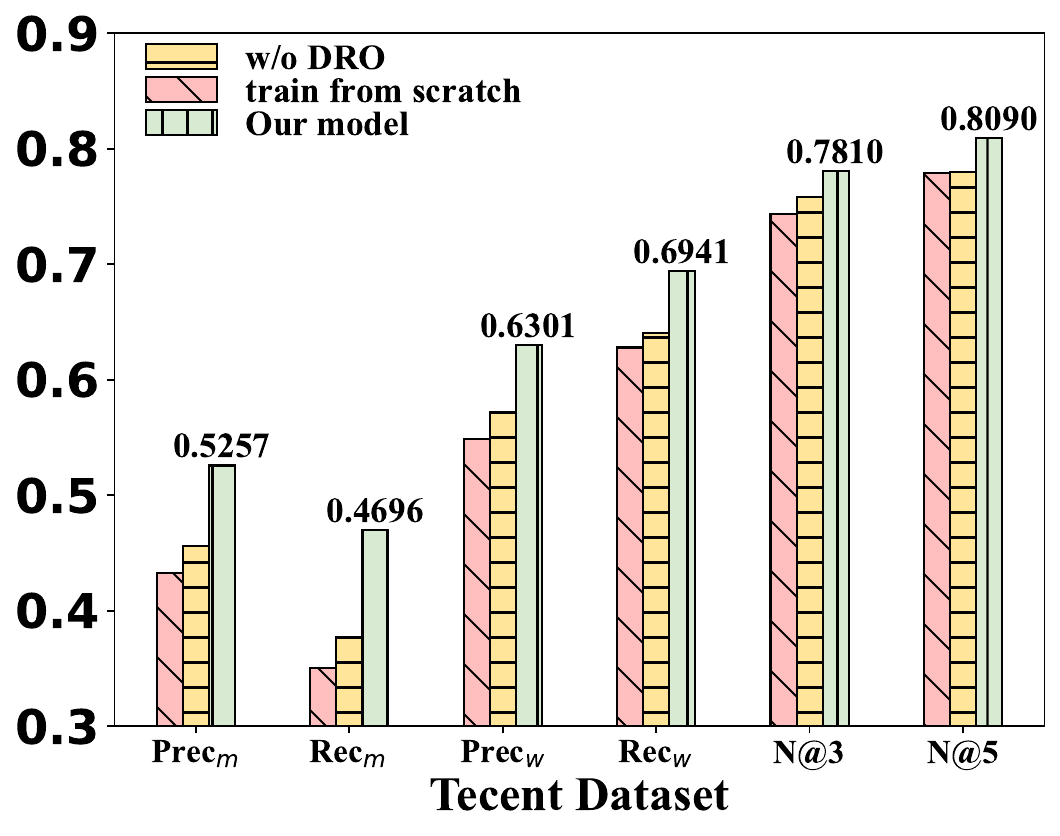}}
\caption{Effectiveness of the DRO-based pretraining paradigm.}
\label{fig:dro}
\end{figure}

\subsection{Effectiveness of the DRO-based Pretraining paradigm} 

The DRO-based pretraining paradigm, enhanced with DRO loss, ensures balanced performance across head and tail classes while improving generalization in the user behavior domain. To evaluate the effectiveness of this paradigm in improving long-tail behavior accuracy, we compared it with long-tail methods such as LLM-ESR, Tailnet, and LIFT. Focusing on five low-proportion behaviors in the Honor dataset, we assessed recall to measure the model's ability to identify long-tail behaviors. Furthermore, we removed the DRO loss and repeated cross-domain experiments to analyze its impact on performance.

As shown in Figure \ref{fig:dro_honor}, our pretraining paradigm outperforms existing long-tail methods in the Honor dataset, effectively balancing performance between mainstream and long-tail behaviors. By addressing the imbalance between head and tail classes, our approach ensures consistent and robust performance across all behavior categories, leading to superior accuracy and recall for both mainstream and long-tail behaviors. 

Additionally, the cross-domain results in Figure \ref{fig:dro_cross} highlight that the DRO loss enhances adaptability, improving generalization and transferability across behavior domains. This ensures high performance across different datasets, further validating the effectiveness of our approach in complex, cross-domain behavior prediction tasks.

\begin{figure}[t]
\centering
\subcaptionbox{Scalibility of BehaveGPT with respective to data size.\label{fig:scaling_data}}{
    \includegraphics[width=0.97\linewidth]{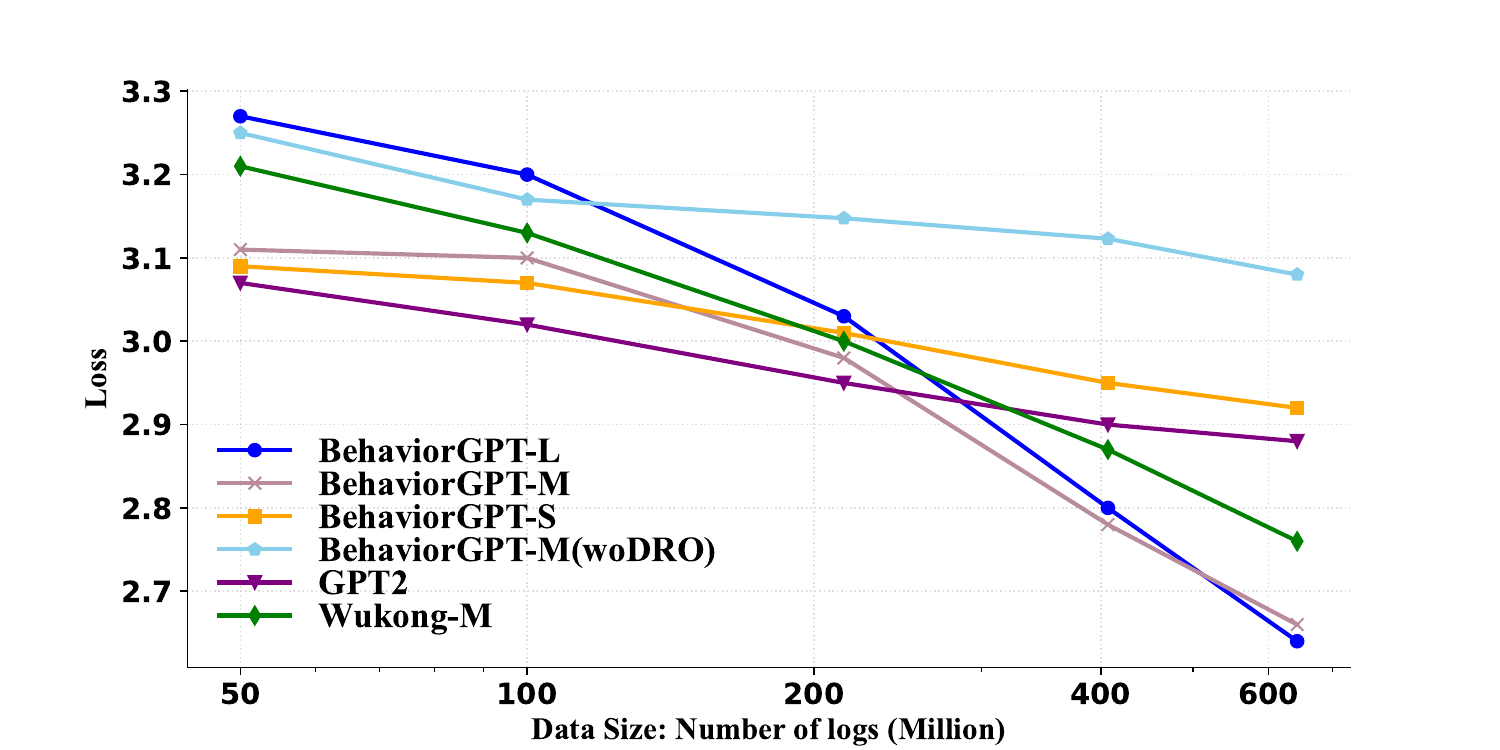}}
\subcaptionbox{Scalibility of BehaveGPT with respective to model parameters.\label{fig:scaling_model}}{
    \includegraphics[width=0.97\linewidth]{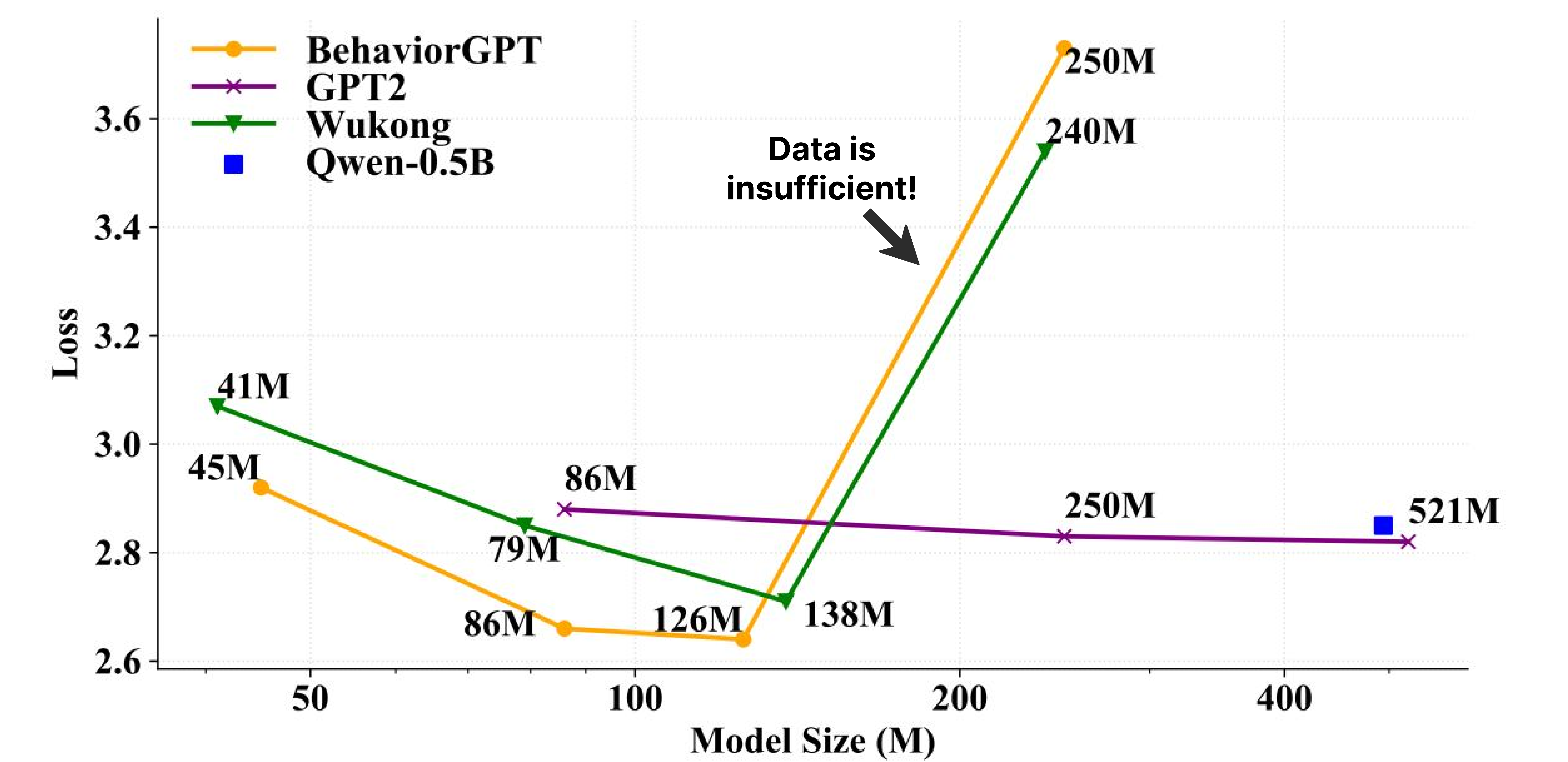}}
\caption{Scalability of BehaveGPT on the Honor datasets.}
\label{fig:scaling}
\end{figure}

\subsection{Scaling Law in user behavior domain}

We aim to analyze the impact of scaling hyperparameters in BehaveGPT on overall model performance. We reference the parameter settings from various versions of the GPT-2 model and train BehaveGPT from scratch across different data scales, with data sizes ranging up to 600 million user behavior records. The results are compared against Wukong \cite{wukong}, a state-of-the-art large-scale recommendation method. To further assess the performance of BehaveGPT relative to LLMs on user behavior data, we also experiment with initializing the model using pretrained LLM parameters.

The scaling phenomenon is shown in Figure \ref{fig:scaling}. From the result, we have the following findings:

\begin{packed_itemize}
\item \textbf{When data is limited, the language-based model outperforms the behavior foundation model.} 
With smaller datasets (e.g., around 50M), the language sequence-based model exhibits lower loss compared to the behavior foundation model. This suggests that, with limited data, the language model benefits from pre-existing knowledge, leading to better performance, while the behavior model struggles due to the constraints of the small training dataset.

\item \textbf{As data size increases, the behavior foundation model becomes more effective. } 
As the dataset expands (e.g., reaching 600M), the behavior foundation model demonstrates a more significant performance improvement, with a sharper decline in loss compared to the language model. This indicates that, with more data, the behavior model is better able to capture user behavior patterns and dependencies, ultimately outperforming the language-based model.

\item \textbf{BehaviorGPT uses data more efficiently compared to other models } 
This is reflected in its lower loss at smaller model sizes. While models like GPT-2 and Wukong require larger datasets to achieve similar performance, BehaviorGPT delivers improved results even with relatively smaller model sizes. This showcases that the DRO-based pretraining paradigm not only balances performance between mainstream and long-tail behaviors but also provides more robust and scalable outcomes overall.
\end{packed_itemize}

Due to data size limitations, the model cannot be scaled indefinitely. Once the model size surpasses a certain threshold, it begins to overfit the training set, reducing its generalization ability and causing a sharp decline in performance. Therefore, exploring the scaling law in the user behavior domain is crucial to understanding the optimal balance between model and data scale. While the scaling law may vary across domains or datasets due to differing data distributions, the optimal data-to-model ratio remains a meaningful metric, offering valuable insights for scaling large models effectively.

We evaluate scaling across both data and model dimensions by training three models, ranging from 0.4M to 24M parameters, each on six different dataset sizes, from $1N$ to $10N$, where $N$ denotes the model size. Then we fit the lossed with model size $N$ and data size $D$ following the curvefit function \cite{hoffmann2022training}:
\begin{equation}
    L(N,D) = C_NN*(-\alpha) + C_DD*(-\beta) +L_0.
\end{equation}

\begin{figure}[t]
\centering
\subcaptionbox{model size: 0.4M.\label{fig:scaling_1}}{
    \includegraphics[width=0.305\linewidth]{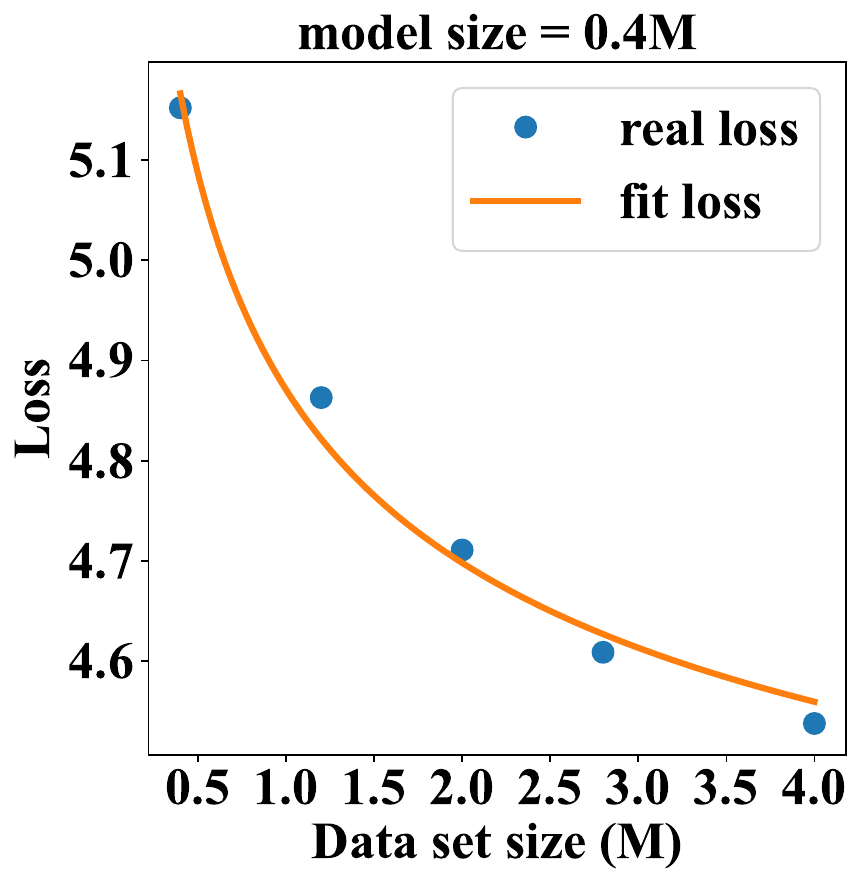}}
\subcaptionbox{model size N: 8M.\label{fig:scaling_2}}{
    \includegraphics[width=0.31\linewidth]{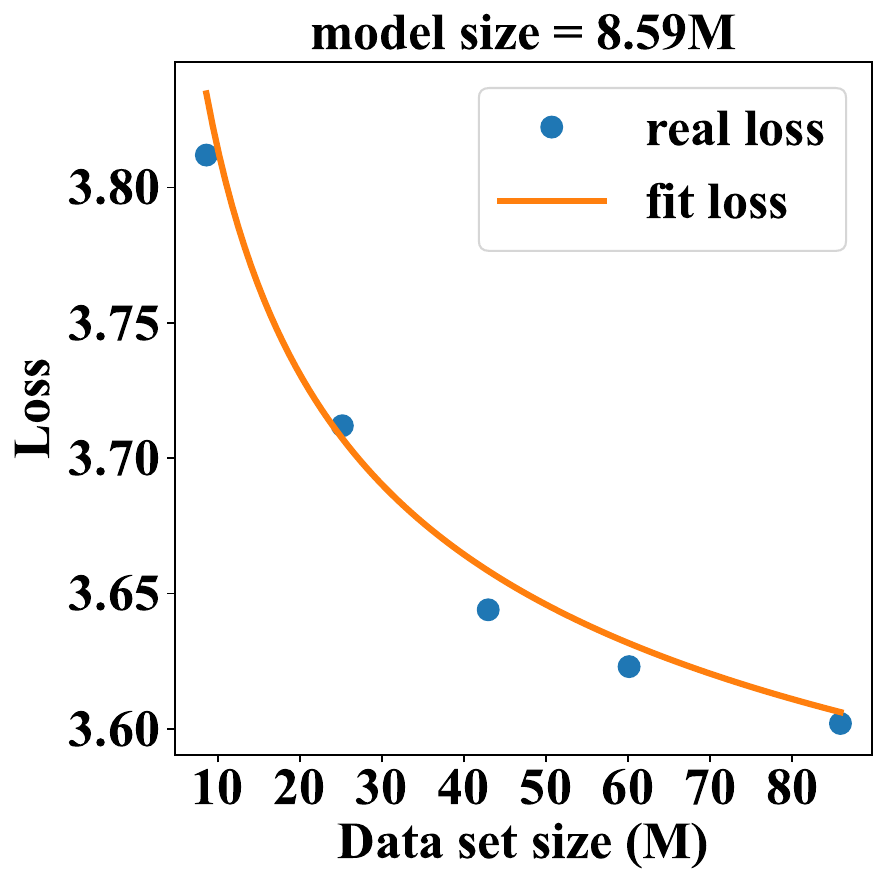}}
\subcaptionbox{model size N: 24M.\label{fig:scaling_3}}{
    \includegraphics[width=0.32\linewidth]{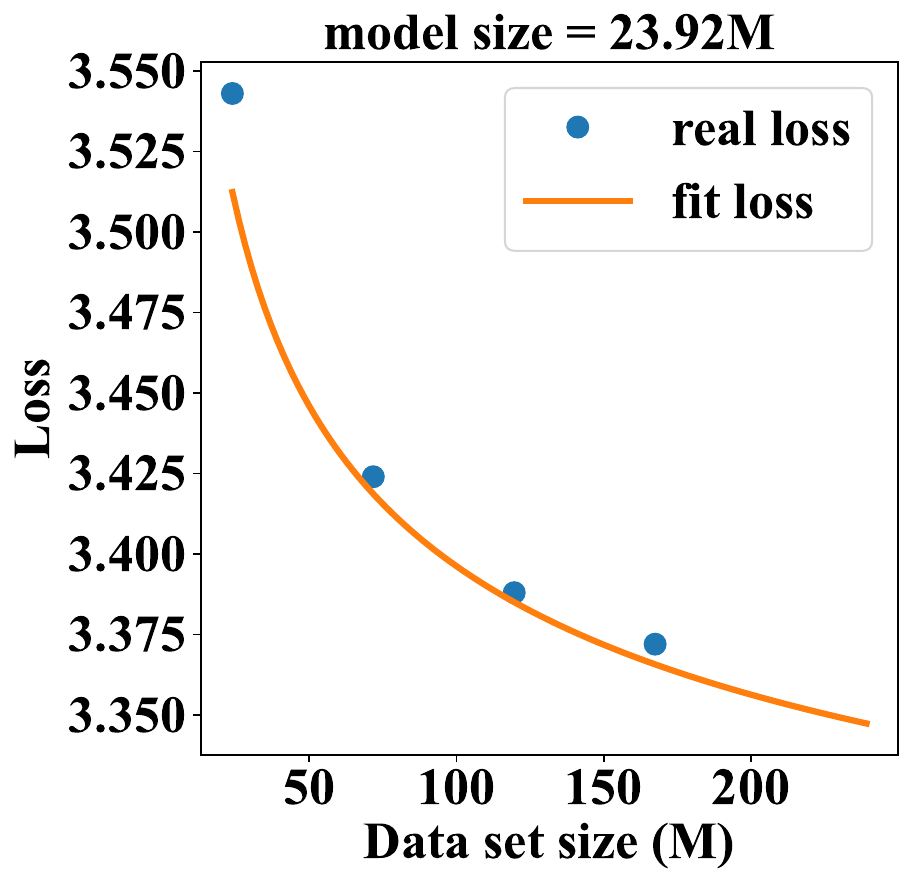}}
\caption{The fitted scaling law plotted along the data size.}
\label{fig:scaling_law}
\end{figure}

In our experiments, the fitted relationship between loss and $N,D$ is shown in Figure \ref{fig:scaling_law}. Specifically, we have $\alpha = 0.51, \beta = 0.23$. Since $\alpha$ is slightly larger than $\beta$, this result shows that as the computation scale, we should slightly emphasize more on data scaling than model scaling. We could also compare the current data-to-model ratio $\frac{D_{opt}}{M_{opt}}$  with that of general LLMs \cite{hu2024minicpm}. Based on the points in Figure 8(a), our data-to-model ratio is approximately 5, while Llama2's ratio ranges from 70 to 100, roughly 15 to 20 times larger than ours. Minicpm's ratio is around 192, about 20 times greater than ours \cite{hu2024minicpm}. This indicates that in the user behavior domain, it is not necessary to use extremely large-scale data to effectively train a foundation model.

\section{Related Works}

\subsection{Foundation Model}
A foundation model is one trained on broad data and adaptable to a wide range of downstream tasks \ cite {li2021foundationmodel}. Their significance can be summarized by two key concepts: emergence and homogenization. Emergence refers to the model's ability to develop new, unanticipated capabilities beyond its explicit programming \cite{emergent}, while homogenization refers to the unification of architectures and training processes \cite{homogenizing}, allowing foundation models to generalize across multiple domains, and reducing the need for specialized, task-specific models.

Foundation models have made their strongest impact in the language domain. BERT \cite{bert}, a bidirectional encoder from Transformers, was the first fine-tuning-based model to achieve state-of-the-art results across multiple benchmarks. GPT \cite{gpt1, gpt2, gpt3}, through self-supervised pretraining and fine-tuning on massive datasets, demonstrated the scaling law, showing that increasing model size and training data progressively enhances performance. LLaMA \cite{llama2} further optimizes with techniques like pre-normalization and the SwiGLU activation function, excelling in tasks related to common sense reasoning and knowledge coverage.

In the CV domain, models like ViT and MAE demonstrate similar scaling principles. ViT \cite{vit} introduces a novel approach by splitting images into patches and using their linear embeddings as input to a transformer. MAE \cite{mae} implements a self-supervised learning method by randomly masking patches of the input image and reconstructing them, paving the way for the expansion of models in CV domain.

Foundation models share key traits: pretraining on large datasets, transferability, scalability, and the ability to develop new capabilities. The transformer architecture, with its self-attention mechanism and scalability,  is the preferred choice for these models. However, due to the complexity and personalization of user behavior,, no foundation model currently exists in this domain.

\subsection{Behavior Modeling}

User behavior modeling emphasizes modeling user-behavior interaction sequences. Recent works \cite{Pu19transformer, Wang2021transformer,ding2020improving} integrate transformers into various models. \citet{yang20iart} introduces intent-aware ranking with transformers, incorporating intent-aware utterance attention. Meanwhile, \citet{wang21masked} proposes a masked-field framework for distinct representations per intent. \citet{quan2023alleviating} propose VLDRec, using multi-task learning to better capture true user preferences. With the rise of large language models, researchers have begun to use LLM agents to simulate user behaviors \cite{gao2023large, yuan2024unist, zhang2025survey}. \citet{yuan2025learning} develops a deep generative collaboration model for mobility behavior generation, which reflects the important spatial-temporal dynamics of human activities. \citet{yuan2024generating} motivated Maslow's need theory, and proposed a knowledge-driven simulation framework based on generative adversarial imitation learning. \citet{peng2025denoising} propose DALR, a denoising alignment framework that bridges GNN and LLM representation spaces by integrating structural and textual features, using contrastive learning to mitigate noise and enhance recommendation performance. \citet{chen2024enhancing} propose LLM4IDRec, a novel framework that leverages large language models to augment ID-only data in recommendation systems, demonstrating that LLMs can effectively enhance ID-based recommendations without relying on textual input.

However, the methods mentioned above cover only a limited range of daily life scenarios, leading to discontinuous and incomplete user behavior data. As a result, they struggle to capture the complex behavioral patterns underlying user actions, making it challenging to evolve these approaches into robust behavior foundation models.

\section{Conclusion}
Inspired by foundational models in NLP and CV, we introduce BehaveGPT, a foundation model designed to understand diverse and dense user behavior data. Utilizing a transformer-based architecture and a novel DRO-based pretraining paradigm, BehaveGPT is trained on large-scale user behavior datasets, enabling it to learn complex patterns and support various downstream tasks such as POI prediction, app usage forecasting, and user behavior prediction.

In our preliminary exploration of scaling law in user behavior modeling, we observe significant improvements in capturing a broader range of user behaviors as the model scales, highlighting its potential as a foundational tool for future research in this domain.

However, BehaveGPT is still an initial effort. The current training datasets do not encompass all possible behavior types, limiting its cross-domain generalization. Future work will focus on collecting more granular behavior data, enhancing adaptation strategies for better transferability across domains, and expanding the range of downstream tasks to further assess its capabilities. By leveraging larger datasets and scaling the model, we aim to increase its versatility and support complex, real-world behavior prediction tasks.

\appendix
\section{Details of baselines}
\label{sec:baseline}
Here we introduce the details of each baseline. For the next behavior prediction task, we have the following 10 baselines: 
\begin{packed_itemize}
\item \textbf{HSTU~\cite{hstu}.}
HSTU reformulates recommendation as sequential transduction problems within a generative framework to handle high cardinality and non-stationary data.

\item \textbf{Wukong~\cite{wukong}.}
Wukong uses stacked factorization machines combined with a synergistic upscaling strategy to establish a scaling law for the recommendation domain.

\item \textbf{LLM-ESR~\cite{llmesr}.}
LLM-ESR enhances sequential recommendation systems by integrating semantic embeddings from LLMs with collaborative signals to address the long-tail item challenge without increasing computational overhead.

\item \textbf{Tailnet~\cite{tailnet}.}
TailNet classifies items into short-head and long-tail categories, introducing a novel preference mechanism to balance user preferences between these two types.

\item \textbf{CDN~\cite{cdn}.}
CDN employs a mixture-of-experts architecture to separate memorization and generalization learning while decoupling user samples from different distributions using a regularized bilateral branch network.

\item \textbf{LIFT~\cite{lift}.}
LIFT emphasizes lightweight fine-tuning over heavy fine-tuning for long-tail learning, enabling fast predictions and compact models through adaptive lightweight strategies.

\item \textbf{PITuning~\cite{pituning}.}
PITuning enhances common pattern extraction and addresses long-tailed preferences using dynamic event-to-intent transition modeling and adaptive unlearning strategies.

\item \textbf{PAT~\cite{pat}.}
PAT redefines the attention mechanism in sequential recommendation by quantifying item dependencies under a global probabilistic model, enhancing diversity.

\item \textbf{CASTR~\cite{castr}.}
CASTR introduces cross-market attention transfer and pretraining strategies to enhance sequential recommendation through selective self-attention transfer and market-specific user preferences.

\item \textbf{SASRec~\cite{sasrec}.}
SASRec uses self-attention to predict the next item in a user's action history by identifying relevant items at each time step.
\end{packed_itemize}

While for the new behavior prediction task, we have the following 2 baselines:
\begin{packed_itemize}
\item \textbf{Mecos~\cite{new_behavior_meta}.}
Mecos, a meta-learning-based cold-start sequential recommendation framework, could extract user preference from limited interactions and learn to match the target cold-start item with the potential user.

\item \textbf{Recformer~\cite{recformer}.}
Recformer combines language understanding and recommendation through a pretraining and fine-tuning methods that flatten item attributes as a  sentence.
\end{packed_itemize}
For the long-term generation task, we have these 3 baselines:
\begin{packed_itemize}
\item \textbf{SeqGAN~\cite{yu2017seqgan}}
SeqGAN uses reinforcement learning with Monte Carlo search for policy updates, avoiding generator differentiation issues.

\item \textbf{DiffuSeq~\cite{gong2022diffuseq}.}
DiffuSeq introduces a diffusion model for sequence-to-sequence tasks, bridging autoregressive and non-autoregressive models.

\item \textbf{UPC-SDG~\cite{liu2022upc}.}
UPC-SDG generates synthetic interaction data based on user privacy preferences, controlling privacy while creating user-specific data.

\end{packed_itemize}

\section{Details of Metrics}
\label{sec:metrics}
We employ the following widely used metrics in prediction task: weighted precision ($Prec_w$), weighted recall ($Rec_w$), macro precision ($Prec_m$), macro recall ($Rec_m$), hit rate (HR) and NDCG(N).

\begin{align} \label{equ:metrics1}
    Prec_w &= \frac{\sum_{b \in B} (\text{TP}_b) \cdot \text{Prec}_b}{\sum_{b \in B} (\text{TP}_b + \text{FP}_b)}, \\ Rec_w &= \frac{\sum_{b \in B} (\text{TP}_b) \cdot \text{Rec}_b}{\sum_{b \in B} (\text{TP}_b + \text{FN}_b)}, \\
    Prec_m &= \frac{1}{|B|} \sum_{b \in B} \frac{\text{TP}_b}{\text{TP}_b + \text{FP}_b}, \\ Rec_m &= \frac{1}{|B|} \sum_{b \in B} \frac{\text{TP}_b}{\text{TP}_b + \text{FN}_b}
\end{align}

Where $|B|$ represents the total number of behaviors. 

While in the long-term generation task, we employ the KS, WD, JSD, BLEU, and Distrinct-2 metrics.
\begin{packed_itemize}

\item \textbf{Kolmogorov-Smirnov (KS)~\cite{daniel1990kstest}}: Assesses the distributional consistency between the predicted and ground-truth data distributions, ensuring the generated outputs align well with the expected behaviors.
\item \textbf{Wasserstein Distance (WD)~\cite{panaretos2019wd}}: Measures the distance between distributions, capturing subtle variations and mismatches between predicted and true behavior patterns.
\item \textbf{Jensen-Shannon Divergence (JSD)~\cite{menendez1997jsd}}: Quantifies the similarity between two probability distributions, providing insights into how closely the model predictions align with the target behavior distributions.
\item \textbf{BLEU~\cite{papineni2002bleu}}: Evaluates the alignment between generated sequences and reference sequences, commonly used to measure the model's precision in capturing target user behaviors.
\item \textbf{Distinct-2~\cite{li2015distinctn}}: Focuses on the diversity of generated outputs, calculating the proportion of unique bigrams in the predictions to ensure behavioral diversity is preserved.
\end{packed_itemize}
These metrics cover key aspects such as distributional consistency, generation diversity, and alignment with target user behaviors.

\bibliographystyle{ACM-Reference-Format}
\balance
\bibliography{Reference}

\appendix

\end{document}